\documentclass[aps,preprint,amssymb,amsmath]{revtex4}
\usepackage[dvipdfm,colorlinks,hypertex]{hyperref}
\usepackage{graphicx}  % Include figure files
\usepackage{bm}
\usepackage{xfrac}
\begin{document}
\title{Exploring the Charge Density Wave phase of 1$T$-TaSe$_2$: Mott or Charge-transfer Gap?}
\author{C. J.~Sayers}
\author{G.~Cerullo}
\affiliation{Dipartimento di Fisica, Politecnico di Milano, Italy}
\author{Y. Zhang}
\author{C. E. Sanders}
\author{R. T. Chapman}
\author{A. S. Wyatt}
\author{G. Chatterjee}
\author{E. Springate}
\affiliation{STFC Central Laser Facility, Research Complex at Harwell, Harwell Campus, Didcot OX11 0QX, UK}
\author{D. Wolverson}
\author{E. Da Como}
\email{edc25@bath.ac.uk}
\affiliation{Centre for Nanoscience and Nanotechnology, Department of Physics, University of Bath, UK}
\author{E.~Carpene}
\email{ettore.carpene@polimi.it}
\affiliation{IFN-CNR, Dipartimento di Fisica, Politecnico di Milano, Italy}
%
%%%%%%%%%%%%%%%%%%%%%%%%%%%%%%%%%%%%%%%%%%%%%%%%%%%%%%%%%%

\begin{abstract}

1$T$-TaSe$_2$ is widely believed to host a Mott metal-insulator transition in the charge density wave (CDW) phase according to the spectroscopic observation of a band gap that extends across all momentum space. Previous investigations inferred that the occurrence of the Mott phase is limited to the surface only of bulk specimens, but recent analysis on thin samples revealed that the Mott-like behavior, observed in the monolayer, is rapidly suppressed with increasing thickness. Here, we report combined time- and angle-resolved photoemission spectroscopy and theoretical investigations of the electronic structure of 1$T$-TaSe$_2$. Our experimental results confirm the existence of a
state above $E_F$, previously ascribed to the upper Hubbard band, and an overall band gap of $\sim 0.7$~eV at $\overline{\Gamma}$. However, supported by density functional theory calculations, we demonstrate that the origin of this state and the gap rests on band structure modifications induced by the CDW phase alone, without the need for Mott correlation effects.

\end{abstract}

% \pacs{78.47.+p}

\maketitle

%%%%%%%%%%%%%%%%%%%%%%%%%%%%%%%%%%%%%%%%%%%%%%%%%%%%%%%%%%

The accurate understanding of electronic phases in quantum materials is a central topic in condensed matter physics. New emergent phenomena often arise from the competition or cooperation of such phases and their complete description can guide the design of material fuctionalities \cite{1,2}. In recent years, charge density waves (CDWs) have been found to coexist in several material classes with other correlated electronic ground states such as superconductivity or Mott-like insulating phases \cite{3,4,5}. The latter involve electron localization in a partially filled band driven by electron-electron interactions and do not necessarily involve changes in the lattice structure or electron-phonon coupling, which are core features of CDWs \cite{6}. When several broken symmetry states of different nature coexist in the same material, it becomes challenging to single out what gives rise to its properties on the macroscopic scale and consequently the response to experimental probes. Low temperature electronic transport has often been used to assess transitions to insulating or superconducting phases, but spectroscopic investigations have deepened our understanding \cite{vo2,vo2b}.

Examples of such materials are the isostructural tantalum-based metal dichalcogenides, 1$T$-TaS$_2$ and 1$T$-TaSe$_2$, well-known CDW systems with several electronic phases \cite{7,8}. The more extensively studied 1$T$-TaS$_2$ exhibits a series of consecutive CDW transitions with increasing commensurability; from incommensurate, to nearly-commensurate to fully commensurate (CCDW) below 180~K. The CCDW transition is characterized by a first-order increase in the resistivity often assigned to a Mott metal-to-insulator transitions (MIT) \cite{9,10}. Mott localization arises from the decreased bandwidth of the half filled Ta-5$d$ band when entering the CDW phase and from the increased on-site electron interactions. In this picture, the band crossing the Fermi level splits into the lower and upper Hubbard bands, LHB and UHB, respectively \cite{11}. 1$T$-TaSe$_2$, which is the focus of this Letter, has a simpler CDW behaviour with a single commensurate transition below 473~K \cite{7}. Thus, it was suggested to be the more ideal compound to investigate the relationship between CDW and Mott \cite{faraday} because of (i) well-separated transition temperatures (CCDW below 473~K, Mott below 260~K \cite{14}), (ii) large electronic gap and (iii) reduced complexity due to absence of non-commensurate CDW phases. The bulk electrical behaviour is consistent with a partially gapped Fermi surface typical of a 2D CDW phase, thus Mott phenomena were not considered in early investigations \cite{9}. However, thanks to the recent growth of single-layer 1$T$-TaSe$_2$ and thus reduced charge screening, the presence of an electronic gap was revealed by angle-resolved photoemission spectroscopy (ARPES) and scanning tunnelling spectroscopy (STS). The nature of this gap was discussed considering Mott electron correlations in density-functional theory (DFT) calculations combined with the Hubbard correction (DFT+U) \cite{12,mottSL2, mottSL3}. Reports of electronic gaps up to $\sim 0.5$~eV in bulk samples by ARPES and STS have been interpreted assuming Mott physics confined to the crystal surface \cite{13,14,15}. Such experimental techniques alone cannot describe the nature of the gap and, while ARPES offers the possibility of a comparison with the calculated band structure, it fails to probe unoccupied bands, which are necessary to estimate gaps across the Fermi surface. Time-resolved ARPES (trARPES) uses an ultrashort pump laser pulse to promote electrons into the conduction band (CB) before the arrival of a probe pulse inducing photoemission, and offers the ability to monitor unoccupied states, their energy dispersion and population dynamics. In addition, the possibility to explore out of equilibrium properties, perturb electronic and lattice order, and launch coherent phonons gives additional information which remains hidden to static techniques \cite{16,17,18}.

In this Letter, we employ trARPES with an infrared pump (0.6~eV) that is resonant with the expected electronic gap of bulk 1$T$-TaSe$_2$ in order to transiently populate the CB and investigate its electron dynamics. We find that the gap evolves on a timescale consistent with electron-phonon interactions rather than that of electron-electron correlations. Using DFT band structure calculations, we further demonstrate how the electronic gap seen in ARPES is in fact a consequence of the CDW reconstruction and does not explicitly require electron-electron correlation effects. In particular, we show that the gap opening at the Fermi level at $\overline{\Gamma}$  is linked to the charge distribution across the CDW unit cell. Our work highlights the relevance of a charge transfer mechanism in CDW systems.

\begin{figure}[htb]
\includegraphics[width=115mm]{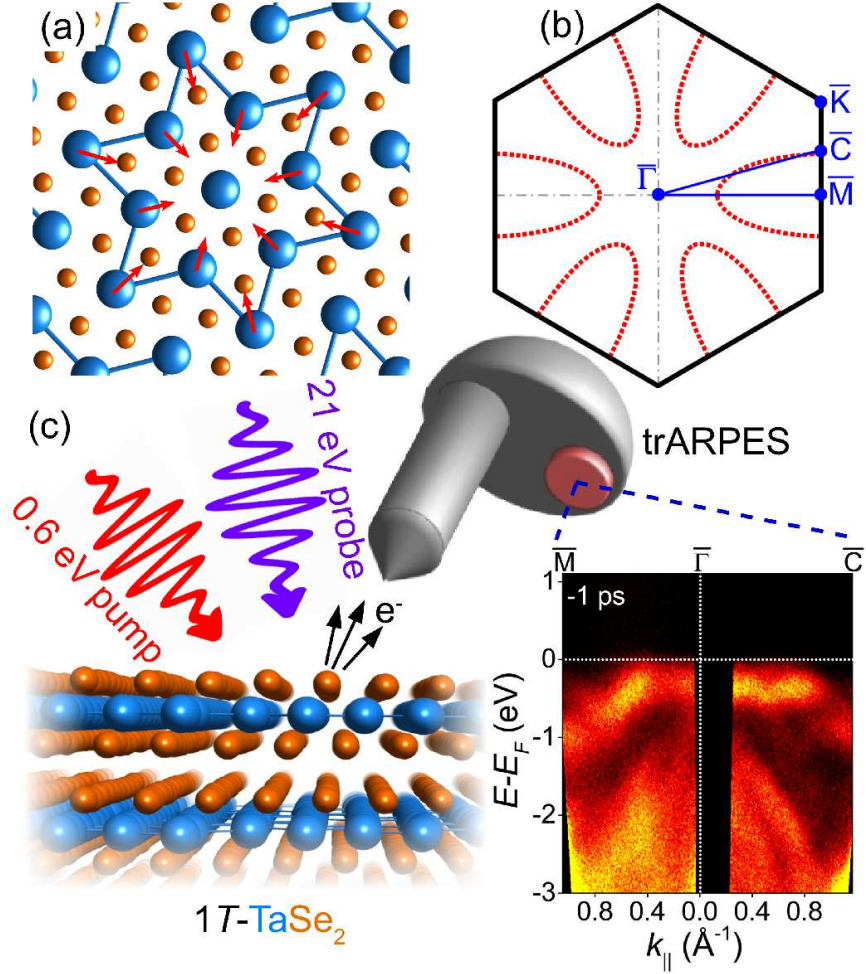}
\caption{%(color online) 
(a) Pictorial view of the star-like lattice reconstruction in the CDW phase of 1$T$-TaSe$_2$. (b) Surface-projected Brillouin Zone (BZ) of the undistorted "normal" state. The red dashed lines mimic the Fermi surface and the blue solid line indicates the experimental path through the BZ as measured by trARPES. (c) Sketch of the trARPES experiment with laser photon energies.}
\label{maps}\end{figure}

The periodic lattice distortion (PLD) of 1$T$-TaSe$_2$ in the charge-ordered phase forms star-like clusters of 13 Ta atoms represented by the blue bonds in Fig.~\ref{maps}a \cite{sod}. The 12 neighboring atoms move towards the center of the star, leading to a $\sqrt{13}\times\sqrt{13}$ reconstruction.
According to previous ARPES investigations \cite{arpes1, arpes2,mott1}, constant binding energy cuts close to the Fermi level reveal large elliptical electron pockets centered at the $\overline{M}$ point of the Brillouin Zone (BZ), as shown in Fig.~\ref{maps}b. Here, we introduce the $\overline{C}$ point which is the midpoint between $\overline{M}$ and $\overline{K}$, and then focus on the $\overline{M\Gamma C}$ path (blue solid line in Fig.~\ref{maps}b) that maximizes the intersection with these pockets. We demonstrate how this is key for observing the bottom CB. Fig.~\ref{maps}c shows the trARPES experimental layout. Photo-excitation was provided by 0.6 eV pulses with fluence of 3~mJ/cm$^2$ at 1~kHz repetition rate. Photoemission spectra were acquired with 21 eV pulses produced via high harmonic generation in argon, which gives access to the entire first BZ of 1$T$-TaSe$_2$. The overall time and energy resolution of the setup is 90~fs and 200~meV, respectively. The sample was held at 80~K during measurements.

\begin{figure}[htb]
\includegraphics[width=115mm]{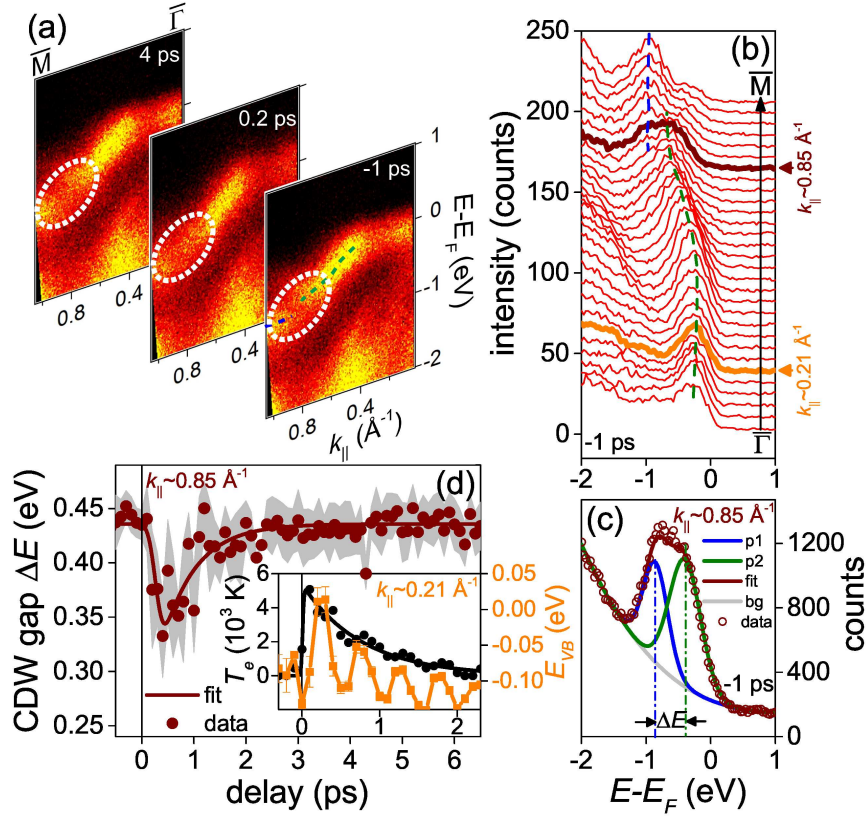}
\caption{%(color online) 
(a) ARPES maps along the $\overline{M\Gamma}$ direction at selected pump-probe delays. Dashed ellipses highlight the location of the CDW gap. (b) EDCs for a range of $k_{||}$ from $\overline{\Gamma}$ to $\overline{M}$ at -1 ps delay. (c) Double peak fitting procedure (see Supplementary C) and (d) temporal evolution of the CDW gap at $k_{||}\simeq 0.85$~\AA$^{-1}$. The grey shaded area is the uncertainty, the brown line is a phenomenological fit. The inset shows the VB dynamics (orange dots, right axis) and the electronic temperature (black dots, left axis) at $k_{||} \simeq 0.21$~\AA$^{-1}$.}
\label{52}\end{figure}

Fig.~\ref{52}a shows the ARPES maps measured along $\overline{M\Gamma}$ at selected pump-probe delays. 
The highest occupied state is the Ta-dominant valence band (VB), which shows a discontinuity, highlighted by dashed ellipses, arising from an avoided crossing due to band folding in the CDW phase (see Supplementary E). At 0.2~ps after the pump (middle map in Fig.~\ref{52}a) the discontinuity is blurred, but it fully recovers within 4~ps. The energy distribution curves (EDCs) at equally separated momenta along $\overline{M\Gamma}$ are shown in Fig.~\ref{52}b for negative pump-probe delay. They reveal a gap at $k_{||}\simeq 0.85$~\AA$^{-1}$ (bold brown curve) whose dynamics can be retrieved through a double-peak fitting procedure, as illustrated in Fig.~\ref{52}c. The peak separation $\Delta E$ represents the CDW gap \cite{faraday} and Fig.~\ref{52}d displays its temporal evolution. It is instructive to compare this with the dynamics observed in the proximity of the Fermi level ($k_{||} \simeq 0.21$~\AA$^{-1}$, bold orange curve in Fig.~\ref{52}b). Upon laser excitation, electrons from occupied states are promptly driven out of equilibrium. Using the Fermi-Dirac function in the EDCs of the VB at $k_{||} \simeq 0.21$~\AA$^{-1}$ (see Supplementary C), the transient electronic temperature is deduced, as reported in the inset of Fig.~\ref{52}d (black dots). The pulsewidth-limited rise time of the electronic temperature is attributed to electron-electron scattering processes in the VB occurring within the pulse duration (sub-$100$~fs). The subsequent cooling is ascribed to electron-phonon interaction with a characteristic scattering time $\tau_{ep}\simeq 0.75$~ps. Additionally, a strong coherent phonon oscillation of the VB is observed (see Fig.~\ref{52}d, orange dots); its frequency $\nu \simeq 2$~THz identifies the so-called breathing mode \cite{KM1,KM2,aom}, that is a coherent expansion and contraction of the star triggered by a displacive mechanism \cite{displacive}, as evidenced by the phase of the oscillations (see Supplementary B). Despite the fact that an intense electronic perturbation is promptly induced by the photoexcitation ($T_e \sim 5000$~K), only a moderate $\sim 22$\% reduction of the CDW gap is observed, with its minimum value being reached about 0.3~ps after the optical excitation (Fig.~\ref{52}d, main panel). Such a response is perfectly compatible with the lattice dynamics, since it matches half the period of the breathing mode ($1/2\nu \simeq 0.25$~ps) and dictates the fastest response time of the lattice to an external perturbation. This fact provides clear evidence that the CDW gap is lattice-related and rather robust against electronic perturbations from an optical excitation. Both CDW gap and electronic temperature recover with the same characteristic time constant $\tau_{ep}$ implying a common mechanism, that is electron-phonon coupling.

\begin{figure}[htb]
\includegraphics[width=115mm]{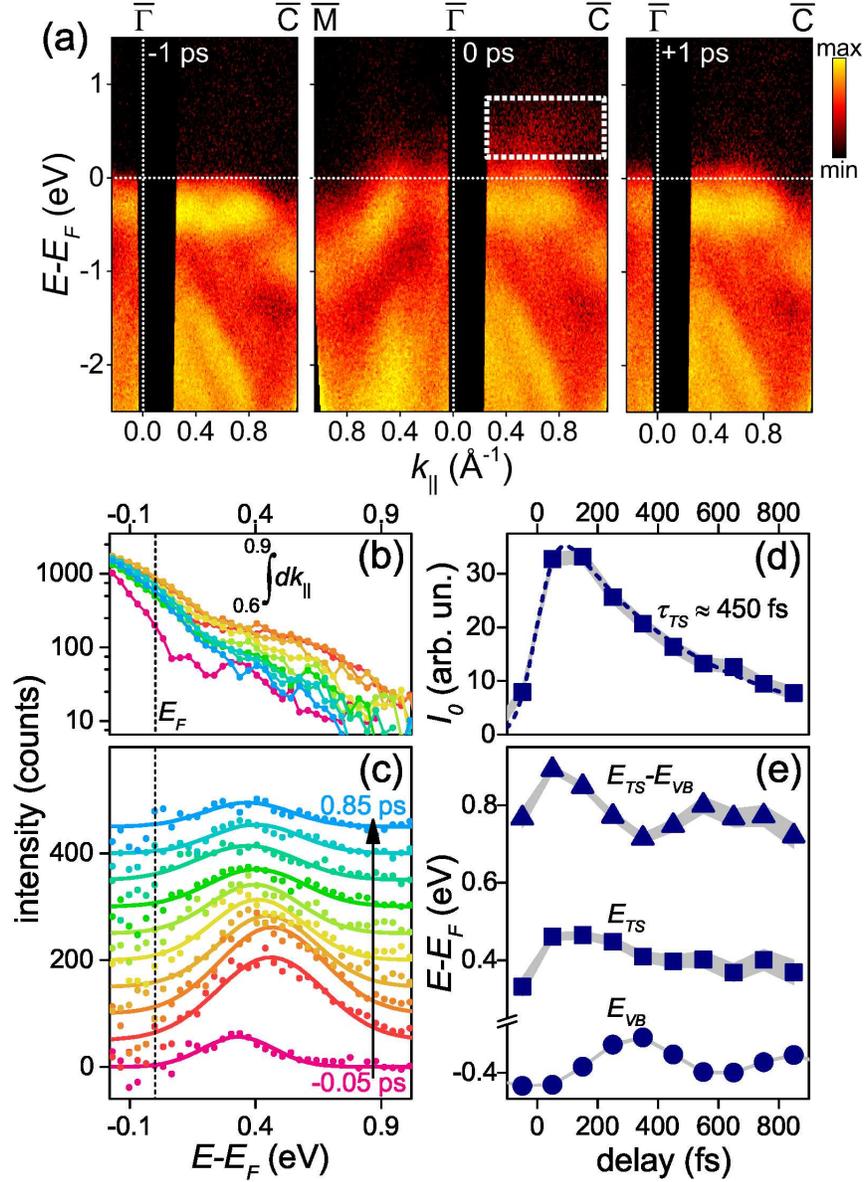}
\caption{%(color online) 
(a) ARPES spectra along the $\overline{(M)\Gamma C}$ path at various pump-probe delays, as indicated. (b) EDCs integrated over $0.6<k_{||}<0.9$~\AA$^{-1}$ ($\overline{\Gamma C}$ direction) at delays ranging from $-0.05$ to 0.85~ps. (c) Same as (b) after highlighting the transiently populated state (TS) above $E_F$. (d) Population dynamics of the TS and (e) binding energy dynamics of TS and VB. The grey shaded areas in panels (d) and (e) are the uncertainties resulting from the peak fitting procedure.}
\label{65}\end{figure}

We now focus on the central experimental result of this Letter. Fig.~\ref{65}a shows the band structure of of 1$T$-TaSe$_2$ along the $\overline{\Gamma C}$ direction as measured by trARPES at selected pump-probe delays (for completeness the full $\overline{M\Gamma C}$ path at 0~ps delay is reported). 
At 0~ps, a transiently populated state (TS) above $E_F$ spanning the probed $\overline{\Gamma C}$ region can be seen (dashed rectangle), which disappears 1~ps after the optical excitation. This state is hardly observed along $\overline{M\Gamma}$. Fig.~\ref{65}b shows the EDCs measured in the time window between $-0.05$ and $+0.85$~ps (step of 0.1~ps) and integrated over $0.6<k_{||}<0.9$~\AA$^{-1}$. After removing the residual spectral weight extending from the VB above the Fermi level, the TS and its dynamics clearly emerge in Fig.~\ref{65}c. The solid lines are Gaussian fits from which intensity and binding energy of the TS are extrapolated, respectively in Fig.~\ref{65}d and Fig.~\ref{65}e. We find a lifetime $\tau_{TS}\simeq450$~fs (Fig.~\ref{65}d) which is shorter than the electron-phonon scattering time, indicating the presence of additional relaxation and recombination channels. The binding energy of the TS (Fig.~\ref{65}e) reveals a weak temporal dependence, converging towards the value $E_{TS}-E_F \simeq0.38$~eV. For comparison, the binding energy $E_{VB}-E_F$ of the VB is reported, showing the effect of the coherent phonon oscillation previously discussed. We notice that the peak-to-peak energy separation between TS and VB, $E_{TS}-E_{VB}$, is essentially modulated by the dynamics of the latter. If we were to label VB and TS as the lower and upper Hubbard bands, we would obtain a Coulomb correlation energy of $\sim0.7$~eV. This value is slightly larger, but still compatible with previous investigations \cite{13,14} within our energy resolution (0.2~eV). However, the lifetime $\tau_{TS}$ of the TS provides a strong indication of its nature.
First, it cannot be ascribed to a laser-induced excitation continuum, as it would bear a much shorter lifetime \cite{ligges} with binding energy rapidly converging towards $E_F$. On the other hand, it cannot be assigned to a Hubbard-like band:
recent trARPES measurements on 1$T$-TaS$_2$ revealed the UHB and its relaxation was estimated to occur on the timescale of electron hopping $\hbar/J\sim 14$~fs \cite{ligges}. Given the structural and electronic similarities with 1$T$-TaS$_2$, one would expect a comparable timescale in 1$T$-TaSe$_2$. However, our data reveal a lifetime of the TS which is $\sim30$ times longer, rebutting its Mott-like nature. DFT calculations also predict the existence of the TS above the Fermi level with a large spectral weight in the $\Gamma C$ direction, as observed in the experiments, but its origin rests on the specific electronic structure induced by the star-like reconstruction, as clarified in the following.

\begin{figure}[htb]
\includegraphics[width=115mm]{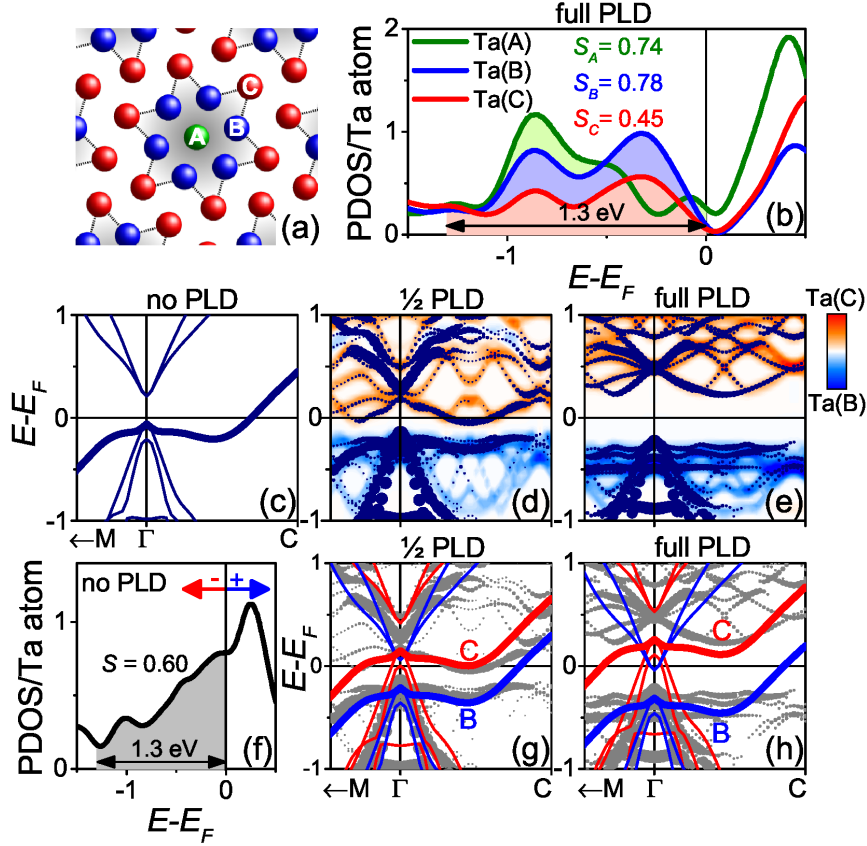}
\caption{%(color online) 
(a) In-plane view of the star-like lattice reconstruction consisting of three different types of Ta atoms. (b) PDOS of each Ta atom type in the CDW phase. Calculated bands of (c) undistorted, (d) 1/2 PLD and (e) full PLD structures along the $M\Gamma C$ path. The dot size in (d)-(e) represents the spectral weigh after band unfolding on the undistorted BZ, while blue- and red-colored bands represent the contributions (projections) of Ta atoms type $B$ and $C$, respectively. (f) Ta PDOS of the undistorted structure. (g)-(h) The Ta-dominant band of the undistorted structure, properly shifted, matches the split bands close to $E_F$ in both $1/2$~PLD and full PLD cases.}
\label{bonds}\end{figure}

In the CDW phase there are three types of nonequivalent Ta atoms labeled $A$, $B$ and $C$ in Fig.~\ref{bonds}a, with numerical proportion $1:6:6$. Atom $A$ (green) lies at the center of the star, atoms $B$ (blue) are the nearest neighbors, atoms $C$ (red) occupy the tips of the star.
The projected density-of-states (PDOS) in Fig.~\ref{bonds}b shows that each atom type has a different electronic occupancy. By integrating the PDOS over the energy range spanned by the highest Ta-dominant VB (from $-1.3$ eV to $E_F$, see Supplementary E), the relative electronic charges $S_{i=A,B,C}$ are estimated: electrons accumulate towards the center of the star ($S_A \simeq S_B\simeq 0.76$), depleting the tips ($S_C \simeq 0.45$) \cite{xps}. This effect, perfectly consistent with the CDW description, is inherently governed by the electronic band configuration in the charge-ordered phase. Focusing our attention on the $M\Gamma C$ path of the BZ, Fig.~\ref{bonds}c reports the calculated band structure of the undistorted 1$T$-TaSe$_2$, i.e. with no PLD. The thick line marks the Ta-dominant band with metallic character (as expected in the normal phase), which would split into LHB and UHB in a Mott insulator picture. Fig.~\ref{bonds}d shows the unfolded band structure at an intermediate distortion (i.e. 1/2 PLD, see Supplementary E) between the undistorted (Fig.~\ref{bonds}c) and the fully-developed CDW phases (Fig.~\ref{bonds}e). Panels (c)-(e) track the band evolution with increasing PLD; the Ta band splits into two main replicas, one moving above the Fermi level and a manifold shifting below it. The former, being partially above $E_F$, hosts fewer electrons than the latter. Based on the charge occupancies inferred from the integrated PDOS (Fig.~\ref{bonds}b), we associate the band below $E_F$ to atoms $A$ and $B$, while its higher energy equivalent to atoms $C$. This scenario is further supported by atom-specific projections: the band structures in Fig.~\ref{bonds}d-e are superimposed to the corresponding contributions of atoms type $B$ (inner part of the star) and type $C$ (tips of the star) in blue and red color, respectively, confirming that the outer portion of the star hosts the unoccupied states, while the inner part spawns the Ta bands below $E_F$.
In the following, we will exploit the calculated PDOS and electronic occupancies to quantitatively estimate the energy shifts of the two Ta-band replicas arising in the charge-ordered phase. Referring to the undistorted structure (Fig.~\ref{bonds}c and ~\ref{bonds}f) one can notice that a rigid shift of the entire band structure to lower[higher] energy with respect to $E_F$ would result in a larger[smaller] electron occupancy (shifting the entire band structure to positive energy corresponds to moving $E_F$ to negative values).
In practice, by moving the Fermi level in Fig.~\ref{bonds}f, we can tune the electronic occupancy $S$ of the Ta-dominant state  in the undistorted lattice to match those of atoms type $B$ or $C$ in the distorted phases. The corresponding shifted bands are shown in Fig.~\ref{bonds}g-h (numerical details for both 1/2 PLD and full PLD cases are given in the Supplementary F). They are superimposed to the respective unfolded band structures (gray dots), revealing a good agreement.
In other words, the VB and the TS experimentally observed by trARPES (Fig.~\ref{65}a) are both Ta-dominant bands originating from different Ta atom types of the star ($B$ and $C$, respectively). Notice that PLD hinders the metallic character, opening a gap across $E_F$. Although the predicted dispersion of the unoccupied Ta band along $\Gamma C$ (Fig.~\ref{bonds}e) cannot be fully probed by our measurements (its bandwidth is comparable to our energy resolution and the pump photon energy of 0.6~eV might not populate it homogeneously) the binding energy determined experimentally matches the computed value well. We point out that our DFT calculations capture all main features of the charge-ordered electronic structure without employing Hubbard correlation terms.
Considering the structural and electronic similarities of 1$T$-TaSe$_2$ and 1$T$-TaS$_2$, one may discuss why the Mott phase appears only in the latter. A comparison between their surface-projected band structures in the undistorted phase (Supplementary G) shows (i) a reduced bandwidth of the Ta-dominant band in the sulfide, as compared to the selenide, indicating stronger electron localization in 1$T$-TaS$_2$ which may favor the Mott transition and (ii) hybridization of chalcogen and metal states at $E_F$ in 1$T$-TaSe$_2$, but not in 1$T$-TaS$_2$, suggesting enhanced screening in the selenide which may hinder the Mott phase.

In conclusion, trARPES data reveal that under photoexcitation conditions leading to a considerable increase of the electronic temperature ($k_B T_e \sim 0.4$~eV, Fig.~\ref{52}d), the induced closure of the CDW gap in the Ta-dominant band is relatively moderate ($\Delta E \sim 0.1$~eV, Fig.~\ref{52}d) and established at the lattice level, as evidenced by the gap response time matching half the period of the star breathing mode ($1/2\nu = 0.25$~ps, Fig.~\ref{52}d, inset). This fact suggests that electron-phonon coupling dominates over electronic instabilities in the formation of the CDW, in accordance with previous works \cite{KM1, insta2, epc1}. The results shown in Fig.~\ref{65} and Fig.~\ref{bonds} prove that, from a purely electronic viewpoint, when entering the CDW phase, charge migration from the tips towards the center of the star leads to the appearance of a {\it charge-transfer} gap rather than a Mott gap \cite{chargetr}, as corroborated by the temporal evolution of the transiently populated state above $E_F$ seen with trARPES, and by DFT calculations. While we cannot disagree with the fact that monolayer 1$T$-TaSe$_2$ might be a Mott system, our investigation confutes the Mott nature of its bulk counterpart.

We acknowledge financial support from the Italian Ministry of University and Research (grant PRIN 2017BZPKSZ) and LaserLab-Europe (grant agreement no. 871124, European Union's Horizon 2020 research and innovation programme). Access to Artemis at the Central Laser Facility was provided by STFC (experiment no. 20120002) with technical support from Alistair Cox, Phil Rice and Ota Michalek. Computational work was also supported by the University of Bath Cloud Pilot Project and the EU Horizon 2020 OCRE project "Cloud funding for research".

%%%%%%%%%%%%%%%%%%%%%%%%%%%%%%%%%%%%%%%%%%%%%%%%%%%%%%%%%%

\end{document}